\documentclass[twocolumn,secnumarabic,amssymb, nobibnotes, aps, prd]{revtex4-2}

\usepackage{amsmath}
\begin{document}

\title{Geometrical interpretation of the wave-pilot theory and manifestation of the spinor fields}%

\author{Mariya Iv. Trukhanova and Gennady Shipov}%
\email[]{trukhanova@physics.msu.ru}
\affiliation{M. V. Lomonosov Moscow State University, Faculty of Physics, Leninskie Gory,  Moscow, Russia}
\date{\today}
\begin{abstract}  Using the hydrodynamical formalism of quantum mechanics for a Schrodinger spinning particle, developed by T. Takabayashi, J. P. Vigier and followers, that involves vortical flows, we propose the new geometrical interpretation of the wave-pilot theory. The spinor wave in this interpretation represents an objectively real field and the evolution of a material particle controlled by the wave is a manifestation of the geometry of space. We assume this field to have a geometrical nature, basing on the idea that the intrinsic angular momentum, the spin, modifies the geometry of the space, which becomes a manifold,  that is represented as a vector bundle with a base formed by the  translational coordinates and time, and  the fiber of the bundle, specified at each point by the field of an tetrad $e^a_{\mu}$, forms from the bilinear combinations of spinor wave function. It was shown, that the spin vector rotates following the geodesic of the space with torsion and the particle moves according to the geometrized guidance equation. This fact explains the self-action of the spinning particle. We show that the curvature and torsion of the spin vector line is determined by the space torsion of the absolute parallelism geometry.
\end{abstract}
\maketitle
\tableofcontents
        \section{Introduction}
        After Schrodinger formulated the famous wave equation in $1926$ \cite{0}, which solution is a de Broglie's plane wave \cite{01}, the issue arose of interpreting the wave function and its physical reality. It is accepted that a wave field, that transfers energy or momentum, can be physically real, as a result, difficulties arose in understanding the physical reality of the wave function. After the Solvay Congress, the so-called Copenhagen or probabilistic interpretation of quantum mechanics was adopted, according to which the wave function was matched by an abstract probability wave. However, not all leading scientists have accepted this interpretation. A. Einstein  argued that the description of quantum phenomena in the framework of the probabilistic concept is not complete, and the deterministic laws are the basis of physical reality.

   In the Louis de Broglie's researches the new deterministic interpretation of the quantum mechanics in terms of continuous and causally determined motion of quantum objects were developed.
After Madelung proposed a hydrodynamic interpretation of quantum mechanics \cite{1}, in which the description based on the Schrodinger wave equation and the complex wave function $\psi=Re^{\frac{i}{\hbar}S}$ was reduced to a description based on two equations for the amplitude $R(\mathbf{r},t)$ and phase $S(\mathbf{r},t)$ of the wave function, Louis de Broglie formulated the fundamental principles of the pilot-wave theory, which was based on three main postulates \cite{2}, \cite{3}, \cite{4}
\begin{itemize}
\item \emph{The particle is localized in space and  moves along a continuous trajectory;}
\item \emph{The wave that is the basis of wave mechanics is physically real object;}
\item \emph{The particle is connected with its wave and, being localized in the wave, moves along one of the streamlines of the hydrodynamical flow, corresponding to the law of wave propagation in the Madelung hydrodynamical model.}
\end{itemize}
Wherein, the internal oscillations of a particle must be in the phase with the propagating wave into which the particle is embedded. Later, in order to establish more fine connection between the particle and the wave, "theory of a double solution" was proposed, in which the concepts of the wave used in the wave mechanics and the real wave field, accompanying a particle, were separated. The statistical wave obeying the Schrodinger wave equation, in this concept, is a probabilistic representation of the state of a particle. The second wave is physically real and unnormalized, and contains a local region of large amplitude and concentration of energy, the singularity, which moves in space. This singularity represents a particle embedded in the wave. In turn, it was predicted, that in the framework of the Madelung's hydrodynamic model, the impact of the surrounding wave on the particle is implemented via quantum potential, which generates a quantum force capable of controlling the motion of the particle \cite{5}, \cite{6}.   Later, the hydrodynamic image of wave propagation and the ideas of the wave-pilot theory were used by David Bohm \cite{5},  who, trying to return to the classical ideal of causality, proposed an interpretation of quantum mechanics in terms of "hidden variables".

On the other hand, in order to build a deterministic interpretation of quantum mechanics, the attention was paid to the geometrical pattern of the representation of interactions. For this purpose, Louis De Broglie attempted to use the analogy between a particle's motion in its wave and the motion of a point mass in a gravitation field \cite{7}, where its trajectory is a geodesic line in the space-time, applying Einstein's reasoning from the general theory of relativity. Another French scientist, Jean-Pierre Vigier, stated the idea to connect the theory of a double solution with the theorem of the motion of a singularity in a gravitational field along geodesic lines of the space-time, which was proved by Einstein together with Grommer.   Vigier developed this analogy, and attempted to represent a particle of matter as a singularity in the space-time metric, surrounded by the wave field.

An important step in constructing the deterministic interpretation was taking a particle with spin into consideration. The causal interpretation of the Pauli equation was initially given using the terms of hydrodynamic model, where the square of modulus of the spinor or wave function  $\psi^*\psi$ determines the fluid density, which has the intrinsic angular momentum connected to the spin. The method used in this article adopts Euler angle representation of the spinor \cite{8}. In parallel to that article, T. Takabayasi obtained  the gauge-independent  tensor formalism for quantum-mechanical non-relativistic particle with spin \cite{9} - \cite{11}, based on the new kind of hydrodynamics of spinning fluid, where each of its element moves like a classical spinning particle under the action of internal spin-potential and internal  magnetic field. In this model, the spinor  is represented in terms of tensors constructed from $\psi^*$ and $\psi$ as bilinear expressions.

Later, T. Takabayasi together with Jean-Pierre Vigier expanded this formalism and represented the picture of the Pauli field as an assembly of very small rotating bodies  continuously distributed in space \cite{9}, using the hydrodynamical model of the Pauli field, where the hydrodynamical fluid carries intrinsic angular momentum to be called "spin". The direction of the angular momentum is fixed to the particle and the angular velocity of rotation points to the direction of special "effective magnetic field" \cite{10}. After that, T. Takabayasi derived the natural extension of the hydrodynamical formalism of the quantum mechanics of a spinning particle in a manner that includes vortical flows \cite{12}. This theory involved geometrical approach and assumed that each element of a hydrodynamical fluid is regarded as a triad structure underlying the classical spin. The theory was constructed in terms of rotation of the triads, which determines the motion of a corpuscle of mass embedded in the spinor wave. This formalism leads to the construction of a realistic physical model of the quantum matter. The main idea of the new interpretation is that the spinor wave represents a new physical field, which exerts an influence on a corpuscle moving within it.

The idea to associate the point-like particle with the singular region of the wave field and use the geometrical approach seems to be very interesting. We follow the hypothesis that the spinor wave is a physical field, that does not have a particular localization in external space-time, but exists as its part. We used the geometrization method based on introduction of the metric of space-time continuum and determination of the geometric tensors, which correspond to the physical fields.

\section{The geometrization the fields of matter}
The electromagnetic, weak and strong interactions are described in the language of gauge theories built on the principle of local internal symmetry, and are not directly related to the space-time geometry. On the other hand, gravity is directly related to the space-time structure for which the local symmetry is the space-time symmetry.

Let's suppose that matter rotation changes the geometry of the event space. Description of the translational motion of physical systems requires introduction of translational coordinates.  However, we assume that when describing rotational motion, the concept of a reference frame based on the laws of light propagation does not fully reflect the properties of space and time.
The hypothesis underlying this paper is that rotation of a physical object generates a field that has new physical nature, and construction of a new theory requires the use of new ideas about the structure of space and time.  From the geometrical point of view, this means that the rotational motion is reflected in a certain way on the geometry of the event space, making it different from the pseudo-Riemannian one \cite{13}, \cite{14}.

In the General Theory of relativity, the spatio-temporal curve, representing the particle motion, is a geodesic. We use the idea of geometrizing the motion of a particle with spin, and represent the field of a spinor wave using concepts borrowed from geometry. Simultaneously, we stop using the concept of a force field, and assign a geometric image to the particle evolution.
\subsection{The field of tetrads}
We follow the idea of the pilot wave theory, according to which the corpuscle of mass is embedded in the wave that guides it. In this case, we may link the center of mass of the particle to the local Lorentzian frame or a tetrad of orthonormal vectors $e_{\mu}^a$.
The simplest generalization of the four-dimensional Minkowskian  geometry to the case of the manifold of oriented points is the geometry of absolute parallelism  \cite{13}, constructed on the  manifold, that is represented as a vector bundle with a base formed by the manifold of the translational coordinates and time. At each point of the base space-time
of the bundle a tangent space or a fibre is specified by the field of a tetrad $e^a_{\mu}$. A base space-time is a $4$-dimensional differential manifold. Space-time coordinates of the base are denoted by $\{x^{\mu}\}$, where Greek $\alpha, \beta, \mu,...$ indices relate to space-time. The tangent space coordinates are be denoted by $\{x^a\}$, where $a, b, c$ determine the tangent space. In general case, the four holonomic coordinates characterize the position of the reference system $e^a_{\mu}$. The dynamical evolution of a tetrad is represented by six coordinates. Six angular coordinates, three of them are spatial angles $\vartheta, \varphi, \chi$ and three of them are the pseudo-Euclidean, that determine the orientation of the tetrad relative to the spatio-temporal axes of the base space-time. The anholonomic coordinates $\{x^a\}$ are given in the bundle layer over a point in the base space and can be introduced as follows  $dx^{\mu}=e^{\mu}_a dx^a$ and $dx^a=e_{\mu}^a dx^{\mu}$, where $e^{\mu}_a\neq\partial x^{\mu}/\partial x_a$.

Let's introduce the orthonormal basis unit vectors of non-rotating laboratory system $\hat{e}^a_{\mu}$ and orthonormal unit basis of  the non-inertial rotating system $e^a_{\mu}$. The basis vectors in a rotating frame of reference are distinguished by a four-dimensional rotation from the basis vectors of a locally-Lorentz inertial frame of reference   \begin{equation}e^{\mu}_a=\Re_a^b\hat{e}^{\mu}_b, \end{equation}  where $\Re_a^b$ represents the matrix of four-dimensional orientation of the position of locally-Lorentz inertial frame of reference relative to its initial position. The matrix of four-dimensional orientation is determined via the product of matrices of pure Lorentz transformations in space-time planes and matrices of three-dimensional rotations. The four-dimensional rotation is decomposed into a pure Lorentz transformation and a three-dimensional rotation around the monoaxis of rotation $\mathbf{n}$ \begin{equation} \Re=R(\mathbf{n})L(\mathbf{v}), \end{equation}  where the three-dimensional speed of the non-inertial reference frame relative to the inertial one is $\mathbf{v}=(v_x, v_y, v_z)$. The matrix of the pure Lorentz transformation is associated with the translational motion of the reference system, and has the form  \cite{140}
\begin{widetext}  \begin{equation}L^b_a=\begin{bmatrix} ch\Theta & -sh\Theta cos\alpha_x & -sh\Theta cos\alpha_y & -sh\Theta cos\alpha_z \\ -sh\Theta cos\alpha_x  & 1+\Delta cos^2\alpha_x & \Delta cos\alpha_x cos\alpha_y & \Delta cos\alpha_x cos\alpha_z \\ -sh\Theta cos\alpha_y & \Delta cos\alpha_x cos\alpha_y & 1+\Delta cos^2\alpha_y & \Delta cos\alpha_x cos\alpha_z \\ -sh\Theta cos\alpha_z & \Delta cos\alpha_x cos\alpha_z & \Delta cos\alpha_y cos\alpha_z & 1+\Delta cos^2\alpha_z \end{bmatrix},\end{equation}\end{widetext} where pseudo-Euclidean angle of rotation of the tetrad $\Theta$ and guide cosines are in the form of  \begin{equation} th\Theta=\frac{|\mathbf{v}|}{c},\qquad \Delta=2sh^2(\Theta/2), \qquad \begin{bmatrix} cos\alpha_x=v_x/v \\ cos\alpha_y=v_y/v \\ cos\alpha_z=v_z/v \end{bmatrix}.\end{equation}    In case of three-dimensional non-relativistic rotation, the rotation matrix can be obtained as a sequence of three turns - clockwise by an angle $\varphi$ around the $\hat{\mathbf{e}}_3$ axis, clockwise by an angle $\vartheta$ around the line of nodes and by an angle $\chi$ around the axis $\mathbf{e}_3$ of the non-inertial frame.
\subsection{Anholonomic space and Torsion}
The tetrad determines the metric tensor of the absolute parallelism geometry
\begin{equation} \label{metric1} g_{\alpha\beta}=\eta_{ab}e_{\alpha}^be^a_{\beta}, \qquad \eta_{ab}= diag (-1, 1, 1, 1) \end{equation}    and the squared distance between two events is $ds^2=g_{\alpha\beta}dx^{\alpha}dx^{\beta}$.
The parallel transfer of the tetrad  relative to the connection $\Delta^{\alpha}_{\beta\gamma}$ is identically equal to zero $\partial_{\gamma}e^a_{\beta}-\Delta^{\alpha}_{\beta\gamma}e^{\beta}_a=0$, so that \begin{equation}\Delta^{\alpha}_{\beta\gamma}=e^{\alpha}_b\partial_{\gamma}e^b_{\beta}. \end{equation}
It was shown by Weitzenbock, that connection of the space with curvature and torsion can be determined in the form of \cite{15}
                     \begin{equation} \label{Connection} \Delta^{\alpha}_{\beta\gamma}=\Gamma^{\alpha}_{\beta\gamma}
                     +T^{\alpha}_{\beta\gamma}, \end{equation}
where \begin{equation}\Gamma^{\alpha}_{\beta\gamma}=\frac{1}{2}g^{\alpha\mu}(\partial_{\gamma}g_{\beta\mu}+\partial_{\beta}g_{\gamma\mu}
-\partial_{\mu}g_{\beta\gamma}) \end{equation} are the Christoffel symbols, and \begin{equation}    T^{\alpha}_{\beta\gamma}=-\Omega^{\ \ \alpha}_{\beta\gamma}+g^{\alpha\mu}\biggl(g_{\beta\nu}\Omega^{\ \ \nu}_{\mu\gamma}+g_{\gamma\nu}\Omega^{\ \ \nu}_{\mu\beta}\biggr)                          \end{equation}            are  the Ricci rotation coefficients, determined via  the object of anholonomity   \cite{17}
             \begin{equation} \label{object}  \Omega^{\ \ \alpha}_{\beta\gamma}=\frac{1}{2}e^{\alpha}_{b}\biggl(\partial_{\beta}e^{b}_{\gamma}-\partial_{\gamma}e^{b}_{\beta}\biggr).
                                                            \end{equation}
  Ricci rotation coefficients can be represented in the form
  \begin{equation}   T^{\alpha}_{\beta\gamma}=e^{\alpha}_b\nabla_{\gamma}e^b_{\beta},\end{equation} where the covariant derivative with respect to the Christoffel symbols is indicated                                                          \begin{equation} \nabla_{\gamma}e^a_{\beta}=\partial_{\gamma}e^a_{\beta}-\Gamma^{\alpha}_{\beta\gamma} e^a_{\alpha},   \end{equation}
 The field of torsion $T^{\alpha}_{\beta\gamma}$ also defines a rotational metric \cite{13}, that describes the rotational properties of an arbitrarily accelerated four-dimensional reference systems
   \begin{equation}  \label{metric2}  d\tau^2=T^{\alpha}_{\beta\gamma}T^{\beta}_{\alpha\mu}dx^{\gamma}dx^{\mu}.                                   \end{equation}
Four equations of motion of the center of mass of the tetrad $x^{\alpha}$ correspond to the translational metric (\ref{metric1})
  \begin{equation}    \frac{d^2x^{\alpha}}{ds^2}+\Gamma^{\alpha}_{\beta\gamma}\frac{dx^{\beta}}{ds}\frac{dx^{\gamma}}{ds}+
 T^{\alpha}_{\beta\gamma}\frac{dx^{\beta}}{ds}\frac{dx^{\gamma}}{ds}=0,     \end{equation}
  while the rotational metric (\ref{metric2}) corresponds to the six rotational equations of motion of the tetrad
   \begin{equation}        \frac{de^{\alpha}_b}{ds}+\Gamma^{\alpha}_{\beta\gamma}e^{\beta}_b\frac{dx^{\gamma}}{ds}+
   T^{\alpha}_{\beta\gamma}e^{\beta}_b\frac{dx^{\gamma}}{ds}=0.             \end{equation}
\subsection{Geometrization of the energy-momentum tensor}
 The curvature tensor in the absolute parallelism space is defined via connection in the usual way \cite{171}
   \begin{equation} \label{Curvature} S^{\alpha}_{\ \beta\gamma\eta}=2\triangle^{\alpha}_{\ \beta[\eta,\gamma]}+2\triangle^{\alpha}_{\ \mu[\gamma}\triangle^{\mu}_{|\beta|\eta]}=0, \end{equation} where the square brackets $[]$ denote alteration of the respective indices and the index enclosed in $| |$ is not subjected to alteration. If we represent the connection of space as the sum of the Christoffel symbols and the Ricci rotation coefficients (\ref{Connection}), and also use the expression for the Riemann tensor of the absolute geometry parallelism space in the form \cite{13}
     \begin{equation}  \Re^{\alpha}_{\ \beta\gamma\eta}=2\Gamma^{\alpha}_{\ \beta[\eta,\gamma]}+2\Gamma^{\alpha}_{\ \mu[\gamma}\Gamma^{\mu}_{|\beta|\eta]},  \end{equation} the expression for the curvature tensor (\ref{Curvature}) takes the form
    $$ S^{\alpha}_{\ \beta\gamma\eta}=\Re^{\alpha}_{\ \beta\gamma\eta}+2T^{\alpha}_{\ \beta[\eta,\gamma]}+2T^{\alpha}_{\ \mu[\gamma}T^{\mu}_{|\beta|\eta]}$$\begin{equation} \label{Curvature 2}+2\Gamma^{\mu}_{\beta[\gamma}T^{\alpha}_{|\mu|\eta]}+2\Gamma^{\alpha}_{\ \mu[\gamma}T^{\mu}_{|\beta|\eta]}=0.  \end{equation} If we add to the right side of the relation (\ref{Curvature 2}) an expression  $-2\Gamma^{\mu}_{[\gamma\eta]}T^{\alpha}_{\mu\beta}=0$ and perform some simple mathematical manipulations, the Riemann tensor can be represented as
    \begin{equation} \label{Curvature3} \Re^{\alpha}_{\ \beta\gamma\eta}=-2T^{\alpha}_{\ \beta[\eta,\gamma]}-2T^{\alpha}_{\ \mu[\gamma}T^{\mu}_{|\beta|\eta]}. \end{equation}Next, using the contraction in the indices $\alpha$ and $\gamma$ of the expression (\ref{Curvature3}) and the indices $\beta$, $\eta$   leads to the scalar curvature
    \begin{equation} \label{Curvature4} \Re=-2g^{\beta\eta}\biggl(\nabla_{[\alpha}T^{\alpha}_{|\beta|\eta]}+2T^{\alpha}_{\mu[\alpha}T^{\mu}_{|\beta|\eta]}\biggr).      \end{equation}   Using the obtained relations (\ref{Curvature3}) and (\ref{Curvature4}), and forming the Einstein tensor $G_{\alpha\beta}=\Re_{\alpha\beta}-\frac{1}{2}g_{\alpha\beta}\Re$, we can derive the equation similar to Einstein's equation, but with a geometrized energy-momentum tensor \begin{widetext} \begin{equation}   T_{\alpha\beta}=-\frac{2}{\nu}\biggl(\nabla_{[\gamma}T^{\gamma}_{|\alpha|\beta]}+T^{\gamma}_{\mu[\alpha}T^{\mu}_{|\gamma|\beta]}
    -\frac{1}{2}g_{\alpha\beta}g^{\delta\lambda}(\nabla_{[\alpha}T^{\alpha}_{|\delta|\lambda]}+T^{\alpha}_{\mu[\alpha}T^{\mu}_{|\delta|\lambda]}) \biggr).   \end{equation} \end{widetext}  As it can be seen from the deduction given above, torsion determines the  Riemannian curvature only in the case if the total Riemann-Cartan curvature tensor is equal to zero. This is possible only within the framework of geometry with absolute parallelism. Since the Ricci rotation coefficients determine the geometrized energy-momentum tensor, it is definitely the torsion fields that can act as the fields of matter. Thus, it is the fields of matter, that are formed by the torsion of the space of absolute parallelism.  The density of matter, which can be determined by the energy-momentum tensor should also have the geometrized structure  \begin{equation} \rho=\frac{g^{\alpha\beta}T_{\alpha\beta}}{c^2}
    =\frac{2}{c^2\nu}\biggl(\nabla_{[\gamma}T^{\gamma}_{|\alpha|\beta]}+T^{\gamma}_{\mu[\alpha}T^{\mu}_{|\gamma|\beta]}\biggr).
      \end{equation}
\section{The geometrical interpretation of wave-pilot}
\subsection{A causal interpretation in the hydrodynamical description}
When constructing the causal interpretation of the Pauli equation, two main ideas were proposed. At first, it was assumed that the spinor wave function of the quantum mechanics $\psi(\mathbf{x},t)$ represents the new kind of physical field and at second, that the square of the wave function $|\psi(\mathbf{x},t)|^2$ represents the density of a fluid, consisting of an ensemble of spinning particles  \cite{8} - \cite{12}. It was postulated that the particle-like inhomogeneity in the fluid or a corpuscle of mass is embedded in the wave and moves with the local stream velocity of the fluid \cite{16}. In this case, the spatial distribution of the ensemble or the fluid density has the form\begin{equation} \label{n} \rho(\mathbf{x},t)=\psi^{\dagger}(\mathbf{x},t)\psi(\mathbf{x},t),\end{equation}
      the center of mass of ensemble or the particle-like inhomogeneity velocity field can be derived in the form
     \begin{equation}\label{v}  v^j(\mathbf{x},t)=\frac{\hbar}{2mi}\frac{\psi^{\dagger}\partial^j\psi
     -\psi\partial^j\psi^{\dagger}}{\psi^{\dagger}\psi}. \end{equation} The fluid has the intrinsic angular momentum connected with spin. The local spin vector which defines the axis of rotation is always pointed along the principal axis of the rigid body representing a particle  \cite{10}, \cite{12}   \begin{equation}  \label{S} s^j(\mathbf{x},t)=\frac{\hbar}{2}\frac{\psi^{\dagger}\sigma^j\psi}{\psi^{\dagger}\psi},  \end{equation}  where $\sigma^j$ are the Pauli matrices.

 Specification of the state of rotation can be done in terms of Euler angles $\vartheta(\mathbf{x},t), \varphi(\mathbf{x},t), \chi(\mathbf{x},t)$, which vary from point to point,  and  the unit spinor $\psi(\mathbf{x},t)$, which defines a state of rotation in the spin space, has the following representation in  terms of the Euler angles
                                              \begin{equation}   \label{psi}     \psi=\begin{pmatrix}
                                              R\cos\frac{\vartheta}{2}e^{-\frac{i}{2}(\varphi+\chi)} \\ R\sin\frac{\vartheta}{2}e^{\frac{i}{2}(-\varphi+\chi)}   \end{pmatrix},
                                                 \end{equation} where $R(\mathbf{x},t)$ is real amplitude.
                                                 The wave function obeys the Schrodinger equation
  \begin{equation}  i\hbar\partial_t\psi(\mathbf{x},t)=\hat{H}\psi(\mathbf{x},t)
\end{equation} with the Hamiltonian  \begin{equation} \label{H}  \hat{H}=\frac{(-i\hbar\hat{\nabla}-\frac{e}{c}\textbf{A})^2}{2m}-\mu_B\hat{\sigma}_{j}B^j, \qquad \mu_B=\frac{e\hbar}{2mc}, \end{equation}where the first term characterizes the kinetic energy of a particle and the second term is the potential energy of the magnetic moment in the external magnetic field $B^j$, $\mathbf{A}$ is the vector potential of external electromagnetic field. Differentiation of particle density with respect to time and application of the Schrodinger equation with Hamiltonian (\ref{H}) lead to continuity equation \begin{equation}    \label{nn}  \partial_t\rho+div(\rho\mathbf{v})=0.        \end{equation}
The spinor is represented in terms of orientation of a spinning body, that makes possible constructing a relationship between the spin rotation, given by Euler angles, and particle velocity \cite{8}
                 \begin{equation}  \label{velocity}   \mathbf{v}=\frac{\hbar}{2m}(\vec{\nabla}\chi+cos\vartheta\vec{\nabla}\varphi)-\frac{e}{c}\mathbf{A}.       \end{equation}
The hydrodynamical representation of the non-relativistic spinning particle is constructed in terms of tensor quantities as the bilinear expressions of $\psi^{\dagger}$ and $\psi$. It is assumed, that the hydrodynamical fluid that carries the intrinsic angular momentum or spin (\ref{S}), has the density distribution (\ref{n}) and velocity distribution (\ref{v}). The quantum effects are represented by non-linear dynamical fields in the equation of motion for the $(\rho, \mathbf{s}, \mathbf{v})$ \cite{16}
\begin{widetext}\begin{equation}    \label{vv} m(\partial_t+\mathbf{v}\cdot\vec{\nabla})\mathbf{v}=\mathbf{F}_L+\frac{e}{mc}s_k\vec{\nabla} B^k+\frac{\hbar^2}{4m}\vec{\nabla}  \biggl( \frac{\triangle n}{n}-\frac{(\nabla n)^2}{2n^2}  \biggr)-\frac{1}{m\rho}\partial_k\biggl(\rho\vec{\nabla}\mathbf{s}\cdot\partial^k \mathbf{s}\biggr),  \end{equation} \begin{equation} \label{ss} (\partial_t+\mathbf{v}\cdot\vec{\nabla})\mathbf{s}=\frac{e}{mc}\mathbf{s}\times\mathbf{B}
+\frac{1}{m\rho}\mathbf{s}\times\partial_k(\rho\partial^k\mathbf{s}).  \end{equation}    \end{widetext}
The external Lorenz force $\mathbf{F}_L$ is represented by the first term on the right side of the equation of motion (\ref{vv}). Each element of the fluid moves under the influence of the quantum Madelung potential, characterized by the third term of the equation (\ref{vv}) and of the "spin stress" represented by the last term. The spin vector precesses due to the influence of external magnetic field represented by the first term of the spin vector evolution equation (\ref{ss}) and "spin torque" is characterized by the last term of equation (\ref{ss}). The "spin stress" and "spin torque" represent the effect of self-action at the hydrodynamic level of description.
\subsection{Geometrical description of the wave-pilot}
On our way to the geometrization of the spinor wave, we introduce two main hypotheses, which we follow throughout this article.   We see no possibility of constructing a mathematical description of the theory, based on a pseudo-Euclidean or a pseudo-Riemannian event space. Our main idea is that the geometry of the event space is closely related to the orientation and movement of the spin vector.
The tetrad of orthonormal vectors lies along the principal axis of the particle-like inhomogeneity and the third axis is directed along the spin vector. The corpuscle can be represented as an oriented point, where the field of frames $e^i_{(a)}(\mathbf{x},t)$, $i, A=1, 2, 3$ is determined with three vectors \cite{12}
    \begin{equation} \label{triad} \mathbf{e}_{(3)}=\mathbf{s}, \qquad  \mathbf{e}_{(1)}= \frac{\mathbf{M}}{\rho}, \qquad \mathbf{e}_{(2)}=\frac{\mathbf{N}}{\rho},  \end{equation} where vectors $\mathbf{M}$ and $\mathbf{N}$ are formed by the combination of the spinor as $\bar{\psi}\sigma^j\psi=M^j+iN^j,$ where $\bar{\psi}^1=-\psi^2$ and $\bar{\psi}^2=\psi^1$.
We consider free motion of a particle without the influence of external fields, where the dynamics of tetrad is directly related to the dynamics of the particle and this relationship has the geometrical nature. Now we should identify the spinor field with changes in the geometry of the event space of absolute parallelism geometry with nonzero torsion. In this case, the motion of the corpuscle should be determined by the geodesic equation in the base coordinates     \begin{equation} \partial_{\gamma}e^{\alpha}_a+\triangle^{\alpha}_{\beta\gamma}e^{\beta}_a=0, \end{equation}
 for the case of a flat space with torsion for the non-relativistic particle  \begin{equation} \label{triad rotation} \partial_ke^i_a+T^i_{jk}e^j_a=0. \end{equation} Using the relationship between the spin vector and the triad (\ref{triad}), as well as the expression for the velocity field via the Euler angles (\ref{velocity}), we derive the geometrized guidance equation, which represents the relationship between the velocity and rotation of the spatial triad, in the geometric picture
      \begin{equation} \label{velocity1} v_i=-\frac{\hbar}{2m}e_{\ (2)}^k\partial_ie^{\ (1)}_k,  \end{equation} where the orientation of the triad satisfies the equation (\ref{triad rotation}). The velocity (\ref{velocity1}) describes twisting around the spin axis $\mathbf{s}$.  Physically, this means that the new physical field is characterized by the course of geodesics in the space of events. Its shape determines the dynamics of the triad, the evolution of which sets the motion of inhomogeneity or, in other words, the triad rotates in a certain way and the inhomogeneity, that transfers the energy, moves.

We can obtain the geometro-hydrodynamical representation of the guidance equation (\ref{velocity1}) in the anholonomic spin space of the bundle, where the triad of vectors is constructed from the bilinear combination of spinor $\psi$.   From the hydrodynamical point of view we can consider the vector line, which configuration is described
by a curve or vector line $s^i, m^i, n^i$, where $m^i$ is a principal normal and $n^i$ is a binormal, taking the position of a point on the
central line as a function of arc length with unit tangent vector $\mathbf{s}$ to the curve.
We can define a Serret-Frenet equation for the triad frame  $\mathbf{e}_a=\mathbf{X}=\{\mathbf{s},\mathbf{m},\mathbf{n}\}$, where $\mathbf{m}=\frac{\mathbf{M}}{\rho}$, $\mathbf{n}=\frac{\mathbf{N}}{\rho}$ \cite{172} \begin{equation}     \frac{\partial\mathbf{s}}{\partial s}=\kappa\mathbf{m},               \end{equation}
 \begin{equation}     \frac{\partial\mathbf{m}}{\partial s}=-\kappa\mathbf{s}+\tau\mathbf{n},               \end{equation}     \begin{equation}     \frac{\partial\mathbf{n}}{\partial s}=-\tau\mathbf{m},               \end{equation}   and the set of other equations for $\mathbf{m}$ and $\mathbf{n}$ direction have the form  \cite{172}
      \begin{equation} \frac{\partial\mathbf{s}}{\partial m}=\theta_{m s}\mathbf{m}+(\Omega_n+\tau)\mathbf{n}, \end{equation}
      \begin{equation} \frac{\partial\mathbf{m}}{\partial m}=-\theta_{m s}\mathbf{s}-(\nabla\cdot\mathbf{n})\mathbf{n}, \end{equation}
      \begin{equation} \frac{\partial\mathbf{n}}{\partial m}=-(\Omega_n+\tau)\mathbf{s}+(\nabla\cdot\mathbf{n})\mathbf{m}, \end{equation}
      and    \begin{equation} \frac{\partial\mathbf{s}}{\partial n}=\theta_{n s}\mathbf{n}-(\Omega_m+\tau)\mathbf{m}, \end{equation}
      \begin{equation} \frac{\partial\mathbf{m}}{\partial n}=(\Omega_m+\tau) \mathbf{s}+(\kappa+(\nabla\cdot\mathbf{m}))\mathbf{n}, \end{equation}
      \begin{equation} \frac{\partial\mathbf{n}}{\partial n}=-\theta_{n s}\mathbf{s}-(\kappa+(\nabla\cdot\mathbf{m}))\mathbf{m}, \end{equation}
where $\tau$ and $\kappa$ are the torsion and curvature of the spin vector line, $\theta_{n s}=\mathbf{n}\cdot\partial_n\mathbf{s}$, $\theta_{m s}=\mathbf{m}\cdot\partial_m\mathbf{s}$, and the divergence of vectors of a moving triad $$ div\mathbf{s}=\theta_{m s}+\theta_{n s}, \qquad div\mathbf{m}=-\kappa+\mathbf{n}\cdot\partial_n\mathbf{m}, $$\begin{equation} div\mathbf{n}=-\mathbf{n}\cdot\partial_m\mathbf{m}. \end{equation}
The spin space is an anholonomic space, where the object of anholonomity is introduced in the form   \begin{equation}   \Omega^a_{bc}=\biggl(\partial_{i}e^a_{k}-\partial_{k}e^a_{i}\biggr)e^{k}_be^{i}_c,   \end{equation}       and the Ricci rotation coefficients $T^a_{bc}=T^a_{ik}e^{k}_be^{i}_c$.
The velocity field in the non-inertial frame can be determined via torsion and curvature of the spin vector line in the following way
    \begin{equation}  \label{velocity2} \mathbf{v}=\frac{\hbar}{2m}\biggl(\tau\mathbf{s}-div(\mathbf{n})\mathbf{m}+(\kappa+div(\mathbf{m}))\mathbf{n}\biggr). \end{equation}
From the geometrical point of view,  the equation of the triad evolution (\ref{triad rotation}) in the context of the Serret-Frenet frame, can be represented in the form   \begin{equation} \label{triad rotation2} \frac{d\mathbf{s}}{ds}=-T^{s}_{jk}\mathbf{e}^j\frac{dx^k}{ds}=\kappa\mathbf{m}, \end{equation} \begin{equation} \label{triad rotation3} \frac{d\mathbf{m}}{ds}=-T^{m}_{jk}\mathbf{e}^j\frac{dx^k}{ds}=-\kappa\mathbf{s}+\tau\mathbf{n}, \end{equation} \begin{equation} \label{triad rotation4} \frac{d\mathbf{n}}{ds}=-T^{n}_{jk}\mathbf{e}^j\frac{dx^k}{ds}=-\tau\mathbf{m}, \end{equation}  where the scalar curvature and torsion of the spin vector line are determined by the Ricci rotation coefficients  $$T^{s}_{mk}\frac{dx^k}{ds}=-T^{m}_{sk}\frac{dx^k}{ds}=-\kappa, $$\begin{equation} T^{m}_{nk}\frac{dx^k}{ds}=-T^{n}_{mk}\frac{dx^k}{ds}=-\tau. \end{equation}
Thus, a particle with a spin moves along a geodesic of space with torsion, which is determined by the object of anholonomity. The curvature and torsion of the spin vector line $s^i$ is determined by the Ricci rotation coefficients. Thus, the self-action of the particle with a spin takes a geometric nature. The tetrad of vectors rotates in a certain way from point to point of the base space in accordance with the geodesic equation (\ref{triad rotation}), and the inhomogeneity moves according to the following law (\ref{velocity2}). According to the Serret-Frenet equations (\ref{triad rotation2}) - (\ref{triad rotation4}) the self-action of the particle is connected with the curvature and torsion of the spin vector line that are determined by the geometry of the space with torsion.

\subsection{Discussion}
Despite the fundamental consilience, the de Broglie-Bohm wave-pilot theory can hardly become generally accepted in its current form, since the physical nature of the spinor wave field is still unknown. Bohm himself already retreated from the concept of a wave-pilot, considering the  quantum non-locality as a subtended characteristic of the space-time, which determines a non-local global field and quantum particle is a vibration mode of this field. Undoubtedly, the wave-pilot theory requires rethinking. We are trying to build a geometric theory in which the evolution of a material particle controlled by a wave is a manifestation of the geometry of space. Based on the results of Ref. \cite{18} we represent the pilot-wave as the manifestation of the geometry of the space, that must be different from the Riemann geometry. In the general form we introduce the manifold of the absolute parallelism geometry, that is represented as a vector bundle with a base formed by the manifold of the translational coordinates and time. At each point of the base space-time
of the bundle there is a tangent space or fiber of the bundle, and the fiber is specified at each point by the field of an tetrad $e^a_{\mu}$. The event space of an arbitrarily accelerated four-dimensional reference system associated with a tetrad is defined by four translational coordinates of the base and six angular coordinates: three spatial angles and three pseudo-Euclidean. Thus, the rotational angles of Euler are embedded in the structure of the space.
We proved that the spin vector of the particle follows the geodesic line of the space with torsion, creating the self-action characterized by the influence of the rotation of the spin on the particle velocity via the guidance equation (\ref{velocity}). On the other hand, the evolution of the triad can be described using the Serret-Frenet equations in the anholonomic spin space. We show, that the curvature and torsion of the spin vector line are determined by the Ricci rotation coefficients of the geometry of absolute parallelism.

The model proposed in the article can help in understanding the basics of the pilot wave concept and fills this theory with a new vision of the nature  of matter and the wave function of quantum mechanics.  The concept of the existence of an empty waves represented by wave functions propagating in space and time but not carrying energy or momentum was proposed by Lucien Hardy and John Stewart Bell \cite{19} - \cite{21}. A. Einstein called this waves - "ghost fields". The question,  may empty waves have  an observable effect, discussed many times \cite{21}, \cite{22}. In the context of the geometrical model of a spinor wave proposed in this article, empty waves can exist independently and transfer information about a material particle that was embedded in the wave.


\begin{thebibliography}{99}
\bibitem{0} E. Schrodinger, Annalen der Physik,  Bd \textbf{79}. S., 361 (1926).
\bibitem{01} Louis de Broglie, Ondeset quanta. Compt. Rend.,  \textbf{179}, 507 (1923).
\bibitem{1}      E. Madelung, Zeit. f. Phys. \textbf{40}, 322 (1926).
\bibitem{2}        L. de Broglie,  J. Physique et le Radium, Issue VI, \textbf{VIII}, 225 (1927); Ann. Fond. Louis de Broglie, \textbf{12}, 1 (1987).
\bibitem{3} L. de Broglie, Tentative d'Interprétation Causale et Non-liniaire de la Micanique
Ondulatoire, Gauthier-Villars, Paris (1956).
\bibitem{4}  L. de Broglie, Les Incertitudes d'Heisenberg et l'Interprétation Probabiliste de la
Mécanique Ondulatoire, Gauthier-Villars and Bordas, Paris (1982).
\bibitem{5} D. Bohm, Phys. Rev., \textbf{85},  180 (1952); \textbf{89}, 458 (1953).
\bibitem{6} D. Bohm and J. P. Vigier, Phys. Rev., \textbf{96}, 208 (1954).
\bibitem{7}  L. de Broglie, J. de Physique. Serie VI., \textbf{8}, 65 (1927).
\bibitem{8} D. Bohm, R. Schiller and J. Tiomno, Suppl. Nuovo Cim., (10), \textbf{1}, 48 (1955).          \bibitem{9} T. Takabayasi and J. P. Vigier, Prog. Theor. Phys., \textbf{18}, 573 (1957).
    \bibitem{10} T. Takabayasi, Prog. Theor. Phys., \textbf{8}, 143 (1952).
    \bibitem{11} T. Takabayasi, Prog. Theor. Phys., \textbf{9}, 187 (1953).
    \bibitem{12} T. Takabayasi, Prog. Theor. Phys., \textbf{70}, 1 (1983).
    \bibitem{13} G. Shipov, A theory of physical vacuum,  Moscow, (1998).
    \bibitem{14} G. Shipov, Russian Physics Journal,  {\bf 10}, 98 (1972).
    \bibitem{140} E. Gubarev, The real principles of relativity, Moscow (2020).
    \bibitem{15} R. Weitzenbock, Proc. Knkl. nederl. akad., {\bf 28}, 400 (1926).
    \bibitem{16} Peter R. Holland, Quantum theory of motion, Cambridge University Press, (2010).
    \bibitem{17} J. A. Schoute, Rieci-Calculus, 2nd ed., Springer, Berlin (1954).
 \bibitem{171} L. Eisenhart, Riemannian geometry. Prinston (N.J.) (1960).
 \bibitem{172} A. W. Marris,   S. L. Passman,  Archive for Rational Mechanics and Analysis, \textbf{32}(1), 29 (1969).
 \bibitem{18}   Mariya Iv. Trukhanova,  Physics Letters A, \textbf{379}, Issue 42, 2777 (2015).
 \bibitem{19} L. Hardy,  Physics Letters A. {\bf167} (1), 11 (1992).
 \bibitem{20}  F. Selleri, A. Van der Merwe,  Quantum paradoxes and physical reality. Kluwer Academic Publishers. p. 85–86, (1990).
 \bibitem{21}   M. Zukowski,   Physics Letters A., \textbf{175} (3–4), 257 (1993).
 \bibitem{22} L. Vaidman,  Foundations of Physics, \textbf{35} (2), 299 (2005).
\end{thebibliography}
    \end{document}